\documentclass[preprint,showpacs,preprintnumbers,amsmath,amssymb]{revtex4}
\usepackage{dcolumn}% Align table columns on decimal point
\usepackage{bm}% bold math
\usepackage{amsfonts}
\usepackage{psfrag}
\usepackage{wrapfig}
\usepackage{subfigure}
\usepackage{makeidx}
\usepackage{multirow}
\usepackage{epsf}
\usepackage{hyperref}
\usepackage{amsmath}

\usepackage{graphicx}
\usepackage{psfrag}
\usepackage{cases}

\makeatletter
\newcommand{\eq}[1]{Eq.(\ref{#1})}
\newcommand{\ud}{\,\mathrm{d}\,}

\newcommand{\Rmnum}[1]{\uppercase\expandafter{\romannumeral #1}}

\begin{document}

\title{$P-V$ criticality of AdS black hole in the Einstein-Maxwell-power-Yang-Mills gravity}

\author{Ming Zhang}
\affiliation{Department of Physics, Northwest University, Xi'an, 710069, China}

\author{Zhan-Ying Yang}
\affiliation{Department of Physics, Northwest University, Xi'an, 710069, China}

\author{De-Cheng Zou}
\email{zoudecheng@sjtu.edu.cn}
\affiliation{Department of Physics and Astronomy, Shanghai Jiao Tong University, Shanghai 200240, China}

\author{Wei Xu}
\affiliation{School of Physics, Nankai University, Tianjin 300071, China\\
School of Physics, Huazhong University of Science and Technology, Wuhan 430074, China}

\author{Rui-Hong Yue}
\affiliation{Department of Physics, Yangzhou University, Yangzhou, 225009, China}

\begin{abstract}

  We study the $P-V$ critical behaivor of N-dimensional AdS black holes in Einstein-Maxwell-power-Yang-Mills gravity.
  Our results show the existence of the Van der Waals like small-large black hole phase transitions
  when taking some special values of charges of the Maxwell and Yang-Mills (YM) fields.
  Further to calculate the critical exponents of the black holes at the critical point, we find that they are the same as those in the Van der Waals liquid-gas system.

\end{abstract}



\maketitle

\section{Introduction}

Thermodynamical properties of the AdS black hole has been a subject of intense study for the past
decades. In terms of the AdS/CFT correspondence, the thermodynamics of black holes in AdS space can be identified
with that of dual strong coupled conformal field theory (CFT) in the boundary of the AdS space \cite{Witten:1998zw}.
The thermodynamic properties in AdS black holes were reported in \cite{Hawking:1982dh},
which represented the existence for a certain phase transition (Hawking-Page phase transition)
in the phase space of the Schwarzschild AdS black hole.

Recently, the analogy between four dimensional RN-AdS black holes and the Van der Waals
fluid-gas system has been completed in this extended phase space \cite{Kubiznak:2012wp},
where the cosmological constant is treated as a thermodynamic pressure with
\begin{eqnarray}
  P=-\frac{\Lambda}{8\pi} \label{equ:p}
\end{eqnarray}
in the geometric units $G_N=\hbar=c=k=1$.
There exist some more meaningful reasons to regard the cosmological
constant as a variable \cite{Kubiznak:2012wp}. Firstly, some more fundamental theories
could be considered, where physical quantities, such as Yukawa coupling,
gauge constant, Newton's constant, or cosmological constant may not be fixed values,
but can vary arising from the vacuum expectation energy. Moreover, the lack of the cosmological
constant $\Lambda$ term in the first law of black hole thermodynamics can not lead to a consistent Smarr
relation for the black hole thermodynamics. In the extended phase space, however, the Smarr relation is satisfied
in addition to the first law of thermodynamics from the aspect of the scaling arguments.
In addition, based on the mathematical analogies,
the $\beta-r_+$ and $Q-\Phi$ diagrams of the fixed charge RN-AdS black holes are found to be
similar to the $P-V$ diagram of the Van der Waals fluid-gas system. Nevertheless, these analogies
are only an identification of similar physical quantities. This analogy becomes more natural
in this extended phase space. Until now, these critical behaviour of a lot of black hole
systems in this extended phase space are under discussion in this direction \cite{Hendi:2012um,Belhaj:2014tga,Xu:2013zea,
Dutta:2013dca,Zou:2014mha, Mo:2014qsa,Wei:2014hba, Liu:2014gvf,Chen:2013ce, Zhao:2013oza,Xu:2014kwa,Cai:2013qga,Xu:2014tja,Ma:2014vxa}.
It is worth to noting that, in four-dimensional Born-Infeld AdS black holes \cite{Gunasekaran:2012dq},
the impact of the nonlinearity can bring the new phenomenon of reentrant phase
transition which was observed in rotating AdS holes \cite{Altamirano:2013ane, Altamirano:2013uqa},
while this reentrant phase transition does not occur for higher dimensional Born-Infeld AdS
black holes \cite{Zou:2013owa}.

So far, the black holes including these two gauge fields (Maxwell field and YM field) are coupled
through gravity have been considered in general relativity \cite{Volkov:1989fi,Brihaye:2006xc,Mazharimousavi:2008ap,
Bostani:2009zf,Bellucci:2011gz,Devecioglu:2014iia,Gao:2003ys}
and higher order derivative gravities \cite{HabibMazharimousavi:2008zz,HabibMazharimousavi:2008ib,Mazharimousavi:2009mb}.
From physics standpoint, electromagnetism has long range effects and dominates outside the nuclei
of natural matter, while YM field is confined to act inside nuclei.
In this paper we will turn to study the black holes in the Einstein-Maxwell-power-Yang-Mills (EMPYM) gravity.
Whether the critical behaviour of black hole with two gauge fields still exist ? It is interesting to explore.
We will find that the existence for the Van der Waals like small-large black hole
phase transition depends on dimension $N$, various values of parameter $q$,
charges $Q$ and $C$ of the YM and Maxwell fields. However, this reentrant phase transition
will not emerge for the black holes with two gauge fields.

This paper is organized as follows. In Sec.~\ref{2s}, we
examine the critical behaviors of the EMPMY black holes. Then,
we will study the critical exponents near critical point in Sec.~\ref{3s}.
By using the Ehrenfest equations, we will evaluate the phase transition of
the EMPMY black holes at the critical point in Sec.~\ref{4s}.
Finally, Sec.~\ref{5s} is devoted to the closing remarks.

\section{Critical behavior of Einstein-Maxwell-power-Yang-Mills black holes}
\label{2s}

The $N$-dimensional action for Einstein-Maxwell-power-Yang-Mills gravity with a cosmological
constant $\Lambda$ is given by
\begin{eqnarray}
{\cal I}=\frac{1}{2}\int dx^N \sqrt{-g}\left(R-\frac{(N-1)(N-2)}{3}\Lambda
-F_{\mu\nu}F^{\mu\nu}-(\textbf{Tr}(F^{(a)}_{\mu\nu}F^{(a)\mu\nu}))^q\right),\label{action}
\end{eqnarray}
where $\textbf{Tr}(.)=\sum_{a=1}^{(N-1)(N-2)/2}(.)$, $R$ is the Ricci scalar, $q$ is a positive real
parameter, YM and Maxwell fields are defined respectively as
\begin{eqnarray}
  &&F_{\mu\nu}^{(a)}=\partial_{\mu}A_{\nu}^{(a)}-\partial_{\nu}A_{\mu}^{(a)}+\frac{1}{2\sigma}
  C_{(b)(c)}^{(a)}A_{\mu}^{b}A_{\nu}^{c}, \\
  &&F_{\mu\nu}=\partial_{\mu}A_{\nu}-\partial_{\nu}A_{\mu}.
\end{eqnarray}
Here $C_{(b)(c)}^{(a)}$ represents the structure constants of $(N-1)(N-2)/2$ parameter
Lie group $G$ and $\sigma$ is a coupling constant, $A_{\mu}^{(a)}$ are the $SO(N-1)$ gauge
group YM potentials, and $A_{\mu}$ is the usual Maxwell potential.

Our metric ansatz for $N$ dimensional spherically symmetric line element is chosen as
\begin{equation}
  ds^2=-f(r)dt^2+\frac{1}{f(r)}dr^2+r^2d\Omega_n^2
\end{equation}
with the line element of a unit $n$-sphere $d\Omega_n^2$. Depending on different dimensions $N$ and
values of $q$, the action Eq.~(\ref{action}) admits various black hole solutions.
As we known, the black hole solutions in the standard Einstein-Maxwell-Yang-Mills theory
with $q=1$ \cite{HabibMazharimousavi:2008zz} and Einstein-power-Yang-Mills theory \cite{Mazharimousavi:2009mb}
have been investigated. Here we present the black hole solutions in the
Einstein-Maxwell-power-Yang-Mills theory, and then discuss the thermodynamics
and critical behavior for each possible black hole solution in the extended phase space.

\subsection{$N(=n+2)\ge 4,~ q\neq(n+1)/4$}
\label{sa}

The solution of $N$-dimensional EMPYM black hole with negative cosmological
constant under the condition of $q\neq(n+1)/4$ is given by
\begin{eqnarray}
f(r)=1-\frac{2m}{r^{n-1}}-\frac{\Lambda}{3}r^2+\frac{2(n-1)C^2}{nr^{2n-2}}+\frac{Q_1}{r^{4q-2}},
\quad Q_1=\frac{[(n-1)nQ^2]^q}{n(4q-n-1)},\label{fr}
\end{eqnarray}
where the integration constant $m$ denotes the mass parameter of black hole,
$C$ and $Q$ are the charges of Maxwell field and Yang-Mills field respectively. Note in order to keep the power Yang-Mils term  satisfying the Weak Energy Condition (WEC), one must take $q>0$ \cite{Mazharimousavi:2009mb}, which is discussed in detailed in what follows. In case of $q=1$, \eq{fr} reduces to the black hole solution in the Einstein-Maxwell-Yang-Mills gravity \cite{HabibMazharimousavi:2008zz}.

In term of the black hole radius $r_+$, the Hawking temperature, mass and entropy
of black hole in the extended phase space read as
\begin{eqnarray}
  &&T=\frac{f'(r_+)}{4\pi}=\frac{n-1}{4\pi r_+}+\frac{2(n+1)P}{3}r_+-\frac{(4q-n-1)Q_1}{4\pi r_+^{4q
  -1}}-\frac{(n-1)^2C^2}{2\pi n r_+^{2n-1}}, \label{equ:t}\\
  &&M=\frac{n\omega_n}{48\pi}\big(8\pi P r_+^{n+1}+3r_+^{n-4q+1}Q_1+3r_+^{n-1}
  +\frac{6(n-1)C^2}{nr_+^{n-1}} \big),\label{equ:M}\\
  && S=\frac{\omega_n r_+^n}{4} \label{equ:s},
\end{eqnarray}
and the YM potential $\Phi_Q$ and the electromagnetic potential $\Phi_C$ can be written as
\begin{eqnarray}
  &&\Phi_Q=\frac{\omega_n q[(n-1)nQ^2]^{q}}{8\pi(4q-n-1)Q}r_+^{n-4q+1}, \\
  &&\Phi_C=\frac{\omega_n (n-1)C}{4\pi r_+^{n-1}},
\end{eqnarray}
where $\omega_n=\frac{2\pi^{(n+1)/2}}{\Gamma(\frac{n+1}{2})}$ is the volume of the unit $n$-sphere.
Moreover, the free energy $F$ of black hole can be written as
\begin{eqnarray}
  F&=&M-T\cdot S.\label{F}
\end{eqnarray}

From Eqs.(\ref{equ:t})(\ref{equ:M})(\ref{equ:s}), these thermodynamic quantities obey the first law of black
hole thermodynamics in the extended phase space
\begin{eqnarray}
  \ud M=T\ud S+\Phi_Q\ud Q+\Phi_C\ud C+V\ud P,
\end{eqnarray}
where $V$ denotes the thermodynamic volume with $V=(\frac{\partial M}{\partial P})_{S,\Phi_Q,\Phi_C}$.
By the scaling argument, we can obtain the generalized Smarr relation for the EMPYM
black hole in the extended phase space
\begin{eqnarray}
  M=\frac{n}{n-1}TS+\Phi_C C+\frac{2q-1}{(n-1)q}\Phi_Q Q-\frac{2}{n-1}VP.
\end{eqnarray}

By rewriting the \eq{equ:t}, we can get the equation of state of the black hole
\begin{eqnarray}
  P=\frac{3T}{2(n+1)r_+}+\frac{3(n-1)^2C^2}{4\pi n(n+1)r_+^{2n}}+\frac{3(4q-n-1)Q_1}{8\pi (n
  +1)r_+^{4q}}-\frac{3(n-1)}{8\pi (n+1)r_+^2}. \label{equ:p0}
\end{eqnarray}
To compare with the Van der Waals fluid equation, we can translate the ``geometric" equation
of state to physical one by identifying the specific volume $v$ of the fluid with the horizon
radius of the black hole as $v=\frac{4r_+}{n}$ such that we will just use the horizon radius in
the equation of state for the black hole hereafter in this paper.

The critical point should satisfy the following condition
\begin{eqnarray}
\frac{\partial P}{\partial r_+}\Big|_{T=T_c, r_+=r_c}
=\frac{\partial^2 P}{\partial r_+^2}\Big|_{T=T_c, r_+=r_c}=0,\label{eq:cp}
\end{eqnarray}
which leads to the critical temperature
\begin{subequations}
  \label{equ:cp}
  \begin{numcases}{}
   T_c=\frac{n-1}{2\pi r_c}-\frac{q(4q-n-1)Q_1}{\pi r_c^{4q-1}}-\frac{(n-1)^2C^2}{\pi r_c^{2n-1}}, \\
   P_c=\frac{3(n-1)}{8\pi(n+1)r_c^2}-\frac{3(4q-1)(4q-n-1)Q_1}{8\pi(n+1)r_c^{4q}}-\frac{3(2n-1)(n-1)^2C^2}{4\pi n(n+1)r_c^{2n}},
  \end{numcases}
\end{subequations}
and the equation for the critical horizon radius
\begin{eqnarray}
  r_c^{4q-2}-2(2n-1)(n-1)C^2 r_c^{4q-2n}-\frac{2q(4q-1)(4q-n-1)Q_1}{n-1}=0. \label{rc}
\end{eqnarray}
The physical solutions of \eq{rc} crucially depend on the dimensions $n$,
parameter $q$ and charges $Q$ and $C$ of the YM and Maxwell fields. Further,
The existence of the critical behavior is also determined by the positive critical pressure
$P_c$ and critical temperature $T_c$.

It is worth to note that \eq{fr} becomes the RN-AdS black hole solution when $Q=0(Q_1=0)$.
The $P-V$ criticality of RN-AdS black hole has been
investigated in \cite{Kubiznak:2012wp,Gunasekaran:2012dq}. Then
we only focus on the cases of $Q\neq0$ below.

\subsubsection{$C=0$}

In the case of $C=0$, \eq{rc} with $q=\frac{1}{2}$ or $q\leq\frac{1}{4}$
does not exhibit any real positive root $r_c$, which implies disappearance of
critical behavior of black hole in these cases. Otherwise \eq{rc} can be solved as
\begin{eqnarray}
  r_c=\left[\frac{2q(4q-1)[(n-1)nQ^2]^q}{n(n-1)} \right]^\frac{1}{4q-2}.\label{r2}
\end{eqnarray}
Plugging it into Eq.(\ref{equ:cp}), the critical temperature and pressure are given by
\begin{subequations}
  \label{equ:cpr2}
  \begin{numcases}{}
   T_c=\frac{(2q-1)(n-1)}{(4q-1)\pi}\left[\frac{2q(4q-1)[(n-1)nQ^2]^q}{n(n-1)} \right]^\frac{-1}{4q-2}, \\
   P_c=\frac{3(n-1)(2q-1)}{16q\pi(n+1)}\left[\frac{2q(4q-1)[(n-1)nQ^2]^q}{n(n-1)} \right]^\frac{-1}{2q-1}.
  \end{numcases}
\end{subequations}
This also gives the following ratio relation
\begin{eqnarray}
  \rho_c=\frac{P_c v_c}{T_c}=\frac{3(4q-1)}{4nq(n+1)},
\end{eqnarray}
where the ratio $\rho_c$ is independent with the charge $Q$ of the YM field.
If taking $q=1$, the system reduces especially to the EYM black hole,
and the critical quantities of EYM black hole are given by
\begin{eqnarray}
r_c=\sqrt{6}Q,\quad T_c=\frac{n-1}{3\sqrt{6}\pi Q},  \quad P_c=\frac{n-1}{32\pi(n+1)Q^2}.
\end{eqnarray}
The ratio here becomes $\rho_c=\frac{P_c v_c}{T_c}=\frac{9}{4n(n+1)}$.

\begin{figure}[htb]
  \subfigure[$q=1$]{\label{fig:subfig:a} %% label for second subfigure
  \includegraphics[width=3.1in]{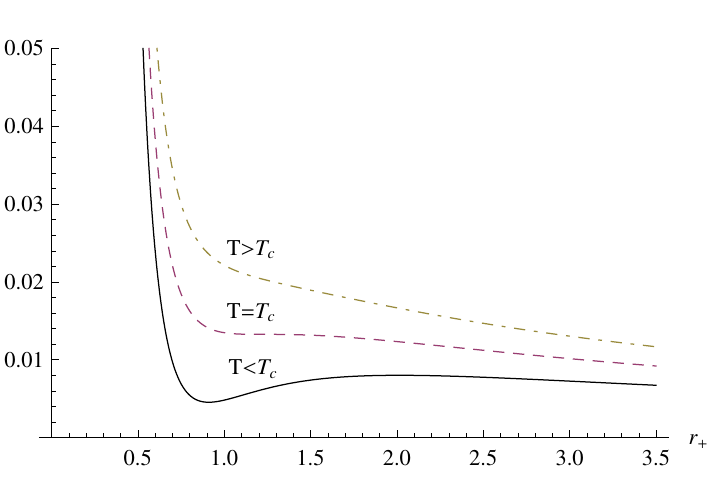}}%
  \hfill%
  \subfigure[$q=2$]{\label{fig:subfig:b} %% label for second subfigure
  \includegraphics[width=3.1in]{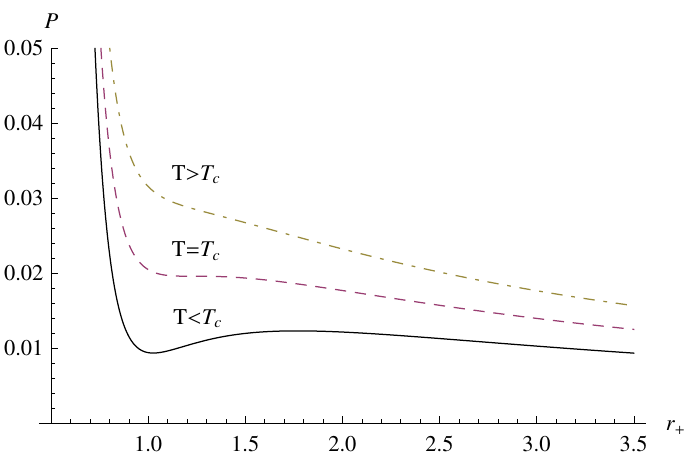}}%
  \caption{ The $P-r_+$ diagrams of the EMPYM black hole with $N=4, Q=0.5, C=0$.
  Here these black, dashed and dotdashed curves correspond to the isotherm $T=0.8T_c$,
  $T=T_c$, $T=1.2T_c$ respectively. }\label{fig1}
\end{figure}

The ``$P-r_+$ diagrams'' of black hole in four dimensions
for $q=1$ and $q\neq1$ are respectively depicted in Fig.\ref{fig1}. These diagrams are exactly the
same as the $P-V$ diagram of the Van der Waals liquid-gas system \cite{Kubiznak:2012wp}.
For a fixed temperature lower than the critical one $T_c$,
we have two branches whose pressure decreases as the increase of horizon radius, one is in the small
radius region (corresponding to fluid phase) and the other is in the
large radius region (corresponding to the gas phase). However, the black holes are always in the
gas phase and no phase transition happens above the critical temperature $T_c$.
In addition, the pictures for free energy shown in Eq.~(\ref{F}) as a function of temperature with
different pressure are plotted in Fig.2.
One can see that these figure develops a ``swallow tail" for $P<P_c$, which denotes a first order
phase transition and this ``swallow tail" vanishes at $P>P_c$ and $P=P_c$ (critical point).
Moreover, the corresponding ``$P-r_+$" and ``$F-T$" diagrams of black hole in higher dimensions
are also qualitatively similar.

%%%%%%%%%%%%%%%%%%%%%%%%%%%%%%%%%%%%%%%%%%%%%%%%%%%%%%%%%%%%%%%%%%%%%%%%%%%

%%%%%%%%%%%%%%%%%%%%%%%%%%%%%%%%%%%%%%%%%%%%%%%%%%%%%%%%%%%%%%%%%%%%%%%%%%%
\begin{figure}[htb]
  \subfigure[$q=1$]{\label{fig:subfig:a1} %% label for second subfigure
  \includegraphics[width=3.1in]{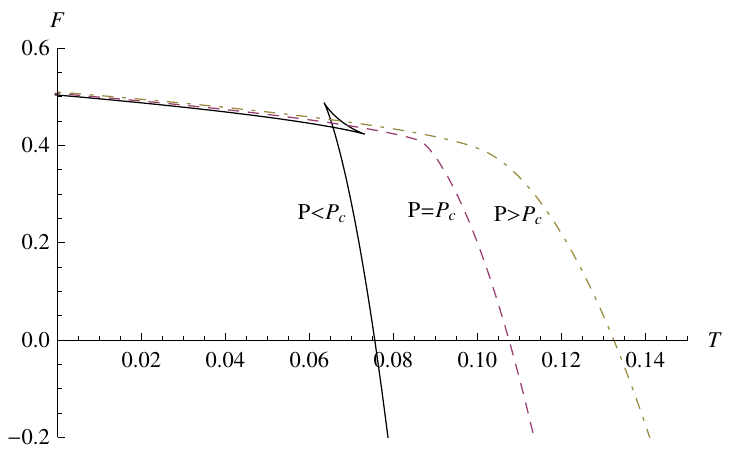}}%
  \hfill%
  \subfigure[$q=2$]{\label{fig:subfig:b1} %% label for second subfigure
  \includegraphics[width=3.1in]{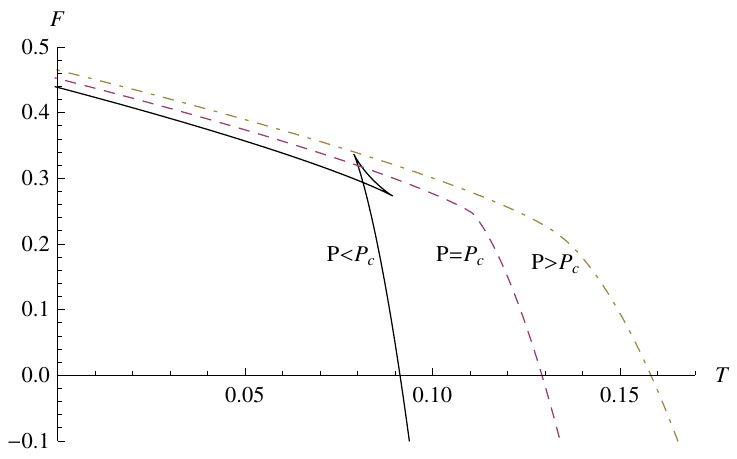}}%
  \caption{ The $F-T$ diagrams of the EMPYM black hole with $N=4, Q=0.5, C=0$.
  Here these black, dashed and dotdashed curves correspond to the isopiestic $P=0.5P_c$,
  $P=P_c$, $P=1.5P_c$ respectively. }\label{fig2}
\end{figure}

\subsubsection{$C\neq 0$}

Generally, it is hard  to exactly solve the \eq{rc} for $C\neq 0$. However, what we are interested in is whether the phase transition happens or not,
which is equivalent to study the existence of positive solution $r_c$ of  \eq{rc}.
We introduce a function
\begin{eqnarray}
H(r_c)&=&r_c^{4q-2}-A r_c^{4q-2n}-B\nonumber\\
&=&r_c^{4q-2}\left(1-\frac{A}{r_c^{2n-2}}\right)-B \label{rs}
\end{eqnarray}
with $A=2(2n-1)(n-1)C^2$ and $B=\frac{2q(4q-1)[(n-1)nQ^2]^q}{n(n-1)}$.
Obviously the critical radius is determined by $H(r_c)=0$.
It is clear that $r_c=A^{1/(2n-2)}$ is a solution if $B=0$ ($q=1/4$).
For generic $q$,  the derivative of $H(r_c)$
\begin{eqnarray}
H'(r_c)=(4q-2)-A\left(4q-2n\right)r_c^{2-2n} \label{rss}
\end{eqnarray}
vanishes at $r_c\equiv r_g=\left[\frac{2(2n-1)(n-1)C^2\left(2q-n\right)}{2q-1}\right]^{\frac{1}{2n-2}}$
 when $q>\frac{n}{2}$ or $q<\frac{1}{2}$, while always maintains positive in the region
of $\frac{1}{2}<q\leq\frac{n}{2}$ and negative for $q=\frac{1}{2}$.

Now, we return to analyse the existence of the critical behaviours based on the values of  $q$.
\begin{enumerate}
  \item $\mathbf{q<\frac{1}{4}}$, the function $H(r_c)$ tends to $-\infty$ as
  $r_c\rightarrow0$ and $H(r_c)\rightarrow(-B)>0$ when $r_c\rightarrow+\infty$, and \eq{rss} has one solution $r_g$.
  This implies that just one positive critical radius  $r_c$ makes  $H(r_c)=0$.
  Meanwhile,  the positive  condition of $T_c,P_c>0$ gives a constrain on  the Maxwell field charge $C$
    \begin{eqnarray}
    \frac{n}{2n-1}\cdot\frac{(1-2q)}{2(n-2q)(n-1)}r_c^{2n-2}<C^2<\frac{(1-2q)}{2(n-2q)(n-1)}r_c^{2n-2}.\label{C}
  \end{eqnarray}

  \item $\mathbf{\frac{1}{4}<q<\frac{1}{2}}$, the function $H(r_c)$ reaches a negative number $(-B)$ as $r_c\rightarrow+\infty$,
  and approaches $-\infty$ if $r_c\rightarrow0$, and \eq{rss} has one solution $r_g$.
  Hence $H(r_c)=0$  admits one or two critical radius solutions $r_c$ which are determined
  by the cases of $H(r_g)=0$ and $H(r_g)>0$. As to the asymptotic behavior of the critical
  temperature $T_c$ (Eq.(18a)) and the critical pressure $P_c$ (Eq.(18b)), we find the temperature function $T_c(r_c)$
  tends to $+\infty$ as $r_c\rightarrow +\infty$, $T_c(r_c)\rightarrow 0$ from the negative direction
  when $r_c\rightarrow 0$ and vanishes at
  \begin{eqnarray}
    r_c\equiv r_t=\left[\frac{2(n-1)\left(n-2q\right)C^2}{1-2q}\right]^{\frac{1}{2n-2}};
  \end{eqnarray}
  the pressure function $P_c(r_c)$ has the same asymptotic behavior with the temperature function and vanished at
  \begin{eqnarray}
    r_c\equiv r_p=\left[\frac{2(n-1)(2n-1)\left(n-2q\right)C^2}{n(1-2q)}\right]^{\frac{1}{2n-2}},
  \end{eqnarray}
  which also gives the inequality $\frac{r_t}{r_g}<\frac{r_p}{r_g}<1$.

  When $H(r_g)=0$, the critical temperature $T_c(r_g)$ is always negative, there is no phase transition. If
   $H(r_g)>0$, the equation $H(r_c)=0$ has two positive solutions $r_1<r_2~$ satisfying  $r_1<r_g<r_2$.
   Taking account of the signs of  $T_c$ and $P_c$ at $r_1$ and $r_2$, one can find that the phase transition happens at $r_1$ under condition

  \begin{eqnarray}
    C^2<\Big[\frac{n(n-1)^2}{2q(n-2q)[(n-1)nQ^2]^q} \Big]^{\frac{n-1}{1-2q}}\cdot\frac{1-2q}{2(n-2q)(n-1)}.
  \end{eqnarray}

  \item $\mathbf{\frac{1}{2}<q<\frac{n}{2}}$, the function $H(r_c)$ reaches positive infinity as $r_c\rightarrow+\infty$ and
  approaches $-\infty$ when $r_c\rightarrow0$.
  Since $H'(r_c)$ is a monotonic increasing function, \eq{rs} only admits just one
  positive critical radius solution $r_c$. Meanwhile, the critical temperature $T_c$ and the critical
  pressure $P_c$ are always positive in this case.

  \item $\mathbf{q>\frac{n}{2}}$, the function $H(r_c)$ approaches $+\infty$ as $r_c\rightarrow+\infty$ and
  tends to a negative number $(-B)$ as $r_c\rightarrow0$. Hence, $H(r_c)=0$ always admits just one
  positive root, and then the charge $C$ satisfies Eq.~(\ref{C}) on account of $T_c,P_c>0$.

  \item $\mathbf{q=\frac{1}{4}}$, critical radius $r_c$ equals to $\big[2(2n-1)(n-1)C^2 \big]^\frac{1}{2n-2}$, and
   the critical temperature and pressure read
  \begin{subequations}
    \label{equ:cpr3}
    \begin{numcases}{}
     T_c=\frac{(n-1)^2}{\pi(2n-1)\big[2(2n-1)(n-1)C^2 \big]^\frac{1}{2n-2}}-\frac{[(n-1)nQ^2]^{1/4}}{4\pi n}, \\
     P_c=\frac{3(n-1)^2}{8\pi n(n+1)\big[2(2n-1)(n-1)C^2 \big]^\frac{1}{n-1}},
    \end{numcases}
  \end{subequations}
  which leads to $C^2<2^{2n-3}\frac{(n-1)^{\frac{7n-9}{2}}n^{\frac{3(n-1)}{2}}}{(2n-1)^{2n-1}Q^{n-1}}$
  based on $T_c>0$ and $P_c>0$.

  \item $\mathbf{q=\frac{1}{2}}$, the solution of \eq{rc} is
  \begin{eqnarray}
    r_c=\big[\frac{2(2n-1)(n-1)C^2}{1-Q_1} \big]^{\frac{1}{2n-2}}.
  \end{eqnarray}
  However, the critical pressure $P_c$ always disappears, there is no phase transition.

  \item $\mathbf{q=\frac{n}{2}}$, the critical quantities can be written as
  \label{equ:cpr4}
  \begin{numcases}{}
    T_c=\frac{(n-1)^2}{\pi(2n-1)r_c}, \\
    P_c=\frac{3(n-1)^2}{8\pi n(n+1)r_c^2}, \\
    r_c=\big[2(2n-1)(n-1)C^2+\frac{(2n-1)[(n-1)nQ^2]^{n/2}}{n-1} \big]^{\frac{1}{2n-2}}.
  \end{numcases}
\end{enumerate}

Therefore, excepting $q=1/2$, there exists a Van der Waals  phase transition under proper conditions, the ``$P-r_+$" and ``$F-T$" diagrams of black holes are similar, which are  depicted in Figs.~3 and 4.
with $N=4$. Moreover, the behaviour of black holes in higher dimensions are also qualitatively similar.

%%%%%%%%%%%%%%%%%%%%%%%%%%%%%%%%%%%%%%%%%%%%%%%%%%%%%%%%%%%%%%%%%%%%%%%%%%%
\begin{figure}[htb]
  \subfigure[$q=1/8$]{\label{fig:subfig:a2} %% label for second subfigure
  \includegraphics[width=3.1in]{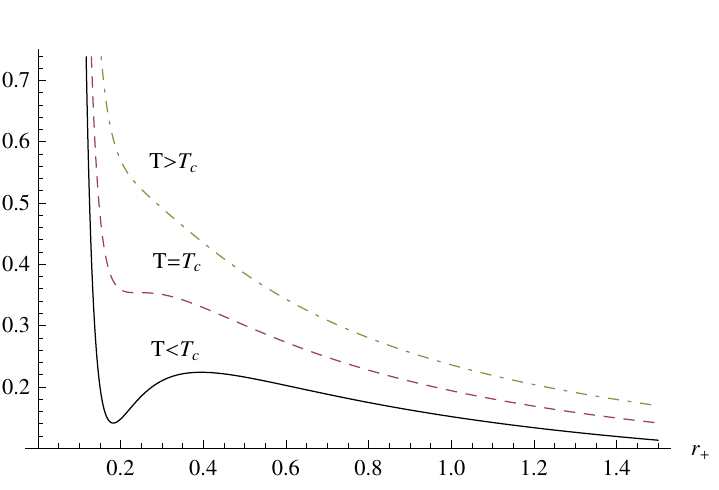}}%
  \hfill%
  \subfigure[$q=1/3$]{\label{fig:subfig:b2} %% label for second subfigure
  \includegraphics[width=3.1in]{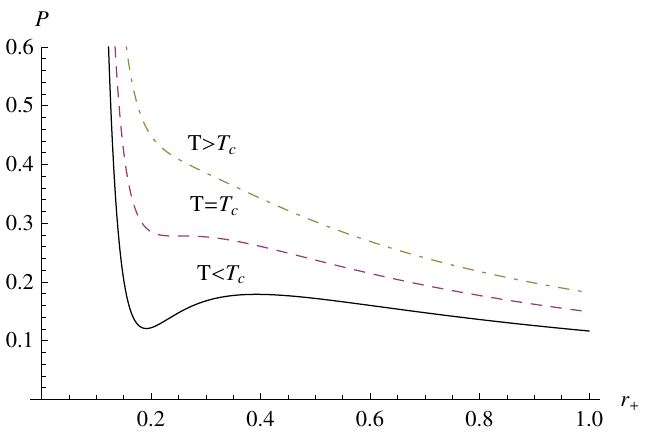}}%
  \caption{ The $P-r_+$ diagrams of the EMPYM black hole with $N=4, Q=1, C=0.1$.
  Here these black, dashed and dotdashed curves correspond to the isotherm $T=0.8T_c$, $T=T_c$, $T=1.2T_c$
   respectively. }\label{fig3}
\end{figure}

%%%%%%%%%%%%%%%%%%%%%%%%%%%%%%%%%%%%%%%%%%%%%%%%%%%%%%%%%%%%%%%%%%%%%%%%%%%
\begin{figure}[h]
  \subfigure[$q=1/8$]{\label{fig:subfig:a3} %% label for second subfigure
  \includegraphics[width=3.1in]{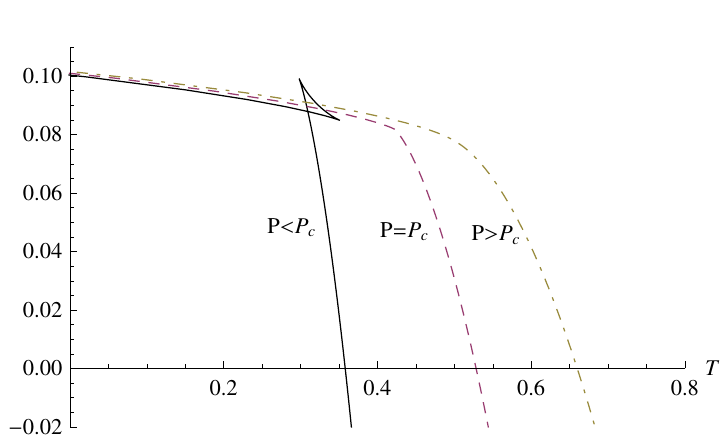}}%
  \hfill%
  \subfigure[$q=1/3$]{\label{fig:subfig:b3} %% label for second subfigure
  \includegraphics[width=3.1in]{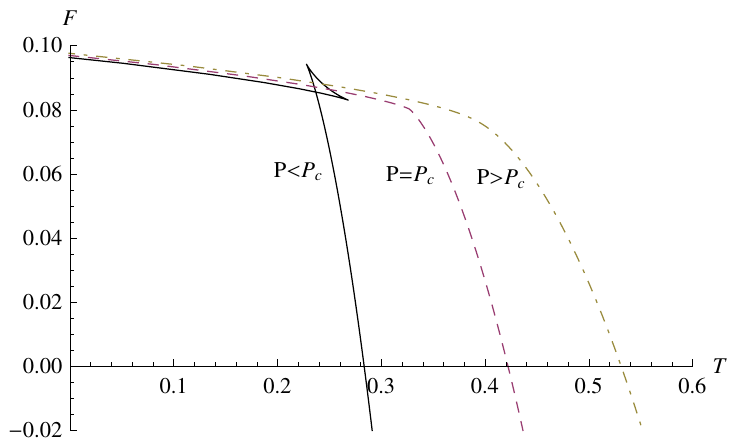}}%
  \caption{ The $F-T$ diagrams of the EMPYM black hole with $N=4, Q=1, C=0.1$.
  Here these black, dashed and dotdashed curves correspond to the isopiestic $P=0.5P_c$, $P=P_c$, $P=1.5P_c$ respectively. }\label{fig4}
\end{figure}

\subsection{$N(=n+2)\ge 4, ~q=(n+1)/4$}
\label{sb}

In this case, the solution of $N$-dimensional EMPYM black hole is given by
\begin{eqnarray}
  &&f(r)=1-\frac{2m}{r^{n-1}}-\frac{\Lambda}{3}r^2+\frac{2(n-1)C^2}{nr^{2n-2}}-\frac{Q_2\ln{r}}{r^{n-1}},\nonumber\\
  &&Q_2=\frac{1}{n}\big[(n-1)nQ^2 \big]^{\frac{n+1}{4}} \label{fr1}.
\end{eqnarray}
When the power exponent $q$ takes 1, \eq{fr1} will reduce to the EMYM black hole where the dimension $n=5$ \cite{HabibMazharimousavi:2008zz}.

The temperature $T$, mass and entropy of this EMPYM black hole in the extended phase space can be derived as
\begin{eqnarray}
 &&T=\frac{n-1}{4\pi r_+}+\frac{2(n+1)Pr_+}{3}-\frac{Q_2}{4\pi r_+^n}-\frac{(n-1)^2C^2}{2\pi n r_+^{2n-1}}, \label{equ:t1}\\
  &&M=\frac{n\omega_n}{8\pi}\left(\frac{r_+^{n-1}}{2}-\frac{\Lambda r_+^{n+1}}{6}-\frac{Q_2\ln{r_+}}{2}+\frac{(n-1)C^2}{nr_+^{n-1}}\right), \\
  &&S=\frac{\omega_n r_+^n}{4}.
\end{eqnarray}
with the pressure $P=-\frac{\Lambda}{8\pi}$. This gives the equation of state
\begin{eqnarray}
  P=\frac{3T}{2(n+1)r_+}+\frac{3Q_2}{8\pi(n+1)r_+^{n+1}}+\frac{3(n-1)^2C^2}{4\pi n(n+1)r_+^{2n}}
  -\frac{3(n-1)}{8\pi(n+1)r_+^2}. \label{equ:p1}
\end{eqnarray}

By adopting the \eq{eq:cp}, we can obtain the critical temperature
\begin{subequations}
  \label{equ:cp1}
  \begin{numcases}{}
   T_c=\frac{(n-1)^2}{2\pi nr_c}+\frac{(n-1)^3C^2}{\pi n r_c^{2n-1}}, \\
   P_c=\frac{3(n-1)^2}{8\pi(n+1)^2r_c^2}+\frac{3(n-1)^3(2n-1)C^2}{4\pi n(n+1)^2 r_c^{2n}}
  \end{numcases}
\end{subequations}
and the equation for the critical horizon radius
\begin{eqnarray}
  r_c^{2n-2}-\frac{n(n+1)Q_2}{2(n-1)}r_c^{n-1}-2(2n-1)(n-1)C^2=0,\label{equ:rc1}
\end{eqnarray}
which leads to
\begin{eqnarray}
  r_c=\left[\frac{n(n+1)Q_2}{4(n-1)}+\sqrt{\frac{n^2(n+1)^2Q_2^2}{16(n-1)^2}+2(2n-1)(n-1)C^2} \right]^{\frac{1}{n-1}}.
\end{eqnarray}
Obviously the critical temperature and pressure are always positive, and then
the critical behaviour also exist in this case. The system reduces to the
EMYM black hole with $q=1$ in five dimensions. Following the same procedure above,
we can also discuss the free energy $F$ of black hole and the first order
phase transition and this ``swallow tail" will also
appear in the $F-T$ diagrams.

\section{Critical exponents near critical point}
\label{3s}

Now we turn to compute the critical exponents $\alpha$, $\beta$, $\gamma$, $\delta$ for the black hole system,
which characterize the behaviors of physical quantities in the vicinity of the critical point $(r_+=r_c, v=v_c, T=T_c, P=P_c)$
for the black hole.
Firstly, we define
\begin{eqnarray}
  p=\frac{P}{P_c};~~\nu=\frac{v}{v_c};~~\tau=\frac{T}{T_c}.
\end{eqnarray}
Near the critical point,  critical exponents are defined as follows \cite{Kubiznak:2012wp}
\begin{eqnarray}
&&C_v=T\frac{\partial S}{\partial T}\Big|_v\propto \left(-\frac{T-T_c}{T_c}\right)^{-\alpha},\nonumber\\
&&\eta=\frac{v_s-v_l}{v_c}\propto \left(-\frac{T-T_c}{T_c}\right)^{\beta},\nonumber\\
&&\kappa_T=-\frac{1}{v}\frac{\partial v}{\partial P}\Big|_T\propto \left(-\frac{T
-T_c}{T_c}\right)^{-\gamma},\nonumber\\
&& P-P_c \propto (v-v_c)^{\delta},\label{eq:27a}
\end{eqnarray}
where $``c"$ denotes the quantity  at the critical point of the system.

In order to compute the critical exponent $\alpha$, we rewrite the entropy of black hole as
$S=\frac{(n+1)\pi^{\frac{n+1}{2}}r_+^n}{4\Gamma(\frac{n+3}{2})}$. Obviously this entropy $S$
is independent of $T$ for the constant value of specific volume $v$, so we conclude that the
critical exponent $\alpha=0$. To obtain the other exponents, we introduce the expansion parameters
\begin{eqnarray}
\tau=t+1, \quad \nu=\omega+1,\label{eq:29a}
\end{eqnarray}
and expand this equation of state near the critical point we can get
\begin{eqnarray}
p=1+a_{10}t+a_{11}t\omega+a_{03}\omega^3+\mathcal{O}(t\epsilon^2,\epsilon^4).\label{eq:30a}
\end{eqnarray}
During the phase transition, the pressure remains constant
\begin{eqnarray}
&&p=1+a_{10}t+a_{11}t\omega_s+a_{03}\omega_s^3=1+a_{10}t+a_{11}t\omega_l+a_{03}\omega_l^3, \nonumber\\
\Rightarrow && a_{11}t\left(\omega_s-\omega_l\right)+a_{03}\left(\omega_s^3-\omega_l^3\right)=0,\label{eq:32a}
\end{eqnarray}
where $\omega_s$ and $\omega_l$ denote the `volume' of small and large black holes.

Using Maxwell's area law, we obtain
\begin{eqnarray}
\int^{\omega_s}_{\omega_l}\omega\frac{dp}{d\omega}d\omega=0 \Rightarrow a_{11}t\left(\omega_s^2-\omega_l^2\right)
+\frac{3}{2}a_{03}\left(\omega_s^4-\omega_l^4\right)=0.\label{eq:33a}
\end{eqnarray}
With Eqs.~(\ref{eq:32a})(\ref{eq:33a}), the nontrivial solutions appear only when $a_{11}a_{03}t<0$. Then we can get
\begin{eqnarray}
\omega_s=\frac{\sqrt{-a_{11}a_{03}t}}{3|a_{03}|}, \quad
\omega_l=-\frac{\sqrt{-a_{11}a_{03}t}}{3|a_{03}|}.\label{eq:34a}
\end{eqnarray}
Table.\ref{tb2} is the different values of $a_{10},~a_{11},~a_{03}$ in \eq{eq:32a} corresponding to
the different situations.
\begin{center}
\begin{table}[h]
  \begin{tabular}{|c|c|c|c|}
  \hline
  \hline
  \scriptsize{label} & $a_{10}$ &  $a_{11}$ &  $a_{03}$  \\
  \hline
  \scriptsize{$Q = 0.5; C = 0; n = 3; q=2$} & 2.29 & -2.29 & -2.67\\
  \hline
  \scriptsize{$Q = 0.5; C = 0.5; n = 3; q=1/4$} & 1.99 & -1.99 & -2.0\\
  \hline
  \scriptsize{$Q = 0.5; C = 0.5; n = 3; q=n/2$} & 2.4 & -2.4 & -2.0\\
  \hline
  \scriptsize{$Q = 0.2; C = 0.4; n = 3; q=(n+1)/4$} & 2.43 & -2.43 & -1.94\\
  \hline
  \scriptsize{$Q = 0.2; C =0.4; n = 5; q=(n+1)/4$} & 2.24 & -2.24 & -3.17 \\
  \hline
  \end{tabular}
  \caption{coefficient} \label{tb2}
\end{table}
\end{center}
Therefore, we have
\begin{eqnarray}
\eta=\omega_s-\omega_l=2\omega_s\propto\sqrt{-t}\Rightarrow \beta=1/2.\label{eq:35a}
\end{eqnarray}

The isothermal compressibility can be computed as
\begin{eqnarray}
\kappa_T=-\frac{1}{v}\frac{\partial v}{\partial P}\Big|_{v_c}\propto
-\frac{1}{\frac{\partial p}{\partial \omega}}\Big|_{\omega=0}=-\frac{1}{a_{11}t},\label{eq:36a}
\end{eqnarray}
which indicates that the critical exponent $\gamma=1$. Moreover, the shape of the critical isotherm $t=0$ is given by
\begin{eqnarray}
p-1=-\omega^3\Rightarrow \delta=3.\label{eq:37a}
\end{eqnarray}
Evidently these critical exponents of the black holes coincide with those of the Van der Waals liquid-gas
system \cite{Kubiznak:2012wp}.

\section{Phase transition at the critical point and Ehrenfest's equations}
\label{4s}

For Van der Waals liquid-gas system, the liquid-gas structure does not change suddenly but undergoes
the second order phase transition at the critical point $(V=V_c, T=T_c, P=P_c)$.
This is described by the Ehrenfest's description \cite{Linder,Stanley}. In conventional thermodynamics,
Ehrenfest's description consists of the first and second Ehrenfest's equations
\cite{sNieuwenhuizen,Zemansky}
\begin{eqnarray}
&&\frac{\partial P}{\partial T}\Big|_S=\frac{C_{P2}-C_{P1}}{TV(\zeta_2-\zeta_1)}
=\frac{\Delta C_P}{TV\Delta\zeta},\label{eq:38a}\\
&&\frac{\partial P}{\partial T}\Big|_V=\frac{\zeta_2-\zeta_1}{\kappa_{T2}-\kappa_{T1}}
=\frac{\Delta\zeta}{\Delta\kappa_{T}}.\label{eq:39a}
\end{eqnarray}
For a genuine second order phase transition, both of these equations have to be satisfied simultaneously.
Here $\zeta$ and $\kappa_T$ denote the volume expansion and isothermal compressibility coefficients
of the system respectively
\begin{eqnarray}
\zeta=\frac{1}{V}\frac{\partial V}{\partial T}\Big|_P,\quad
\kappa_T=-\frac{1}{V}\frac{\partial V}{\partial P}\Big|_T. \label{eq:40a}
\end{eqnarray}

Let us concentrate on the $N$-dimensional EMPYM black hole. From Eq.~(\ref{eq:40a}), we can obtain
\begin{eqnarray}
V\zeta=\frac{\partial V}{\partial T}\Big|_P
=\frac{\partial V}{\partial S}\Big|_P\times\frac{\partial S}{\partial T}\Big|_P
=\frac{\partial V}{\partial S}\Big|_P\times\frac{C_P}{T}.\label{eq:41a}
\end{eqnarray}
The right hand side of Eq.~(\ref{eq:38a}) can be expressed into
\begin{eqnarray}
\frac{\Delta C_P}{TV\Delta \zeta}=\left[\frac{\partial S}{\partial V}\Big|_P\right]_{r_+=r_c}
=\frac{3}{2(n+1)r_c},\label{eq:42a}
\end{eqnarray}
where the thermodynamic volume $V$ is described above and the subscript denotes the physical
quantities at the critical point. From \eq{equ:p0} and \eq{equ:p1},
the left hand side of Eq.~(\ref{eq:38a}) at the critical point can be got as
\begin{eqnarray}
\left[\frac{\partial P}{\partial T}\Big|_S\right]_{r_+=r_c}
=\frac{3}{2(n+1)r_c}.\label{eq:43a}
\end{eqnarray}
Therefore, the first of Ehrenfest's equations can be satisfied at the critical point.

Now let's examine the second of Ehrenfest's equations. In order to compute $\kappa_T$, we
make use of the thermodynamic identify
\begin{eqnarray}
\frac{\partial V}{\partial P}\Big|_T\times\frac{\partial P}{\partial T}\Big|_V
\times\frac{\partial T}{\partial V}\Big|_P=-1.\label{eq:44a}
\end{eqnarray}
Considering Eq.~(\ref{eq:40a}), we can have
\begin{eqnarray}
\kappa_T V=-\frac{\partial V}{\partial P}\Big|_T=\frac{\partial T}{\partial P}\Big|_V
\times\frac{\partial V}{\partial T}\Big|_P
=\frac{\partial T}{\partial P}\Big|_V V\zeta.\label{eq:45a}
\end{eqnarray}
which reveals the validity of the second Ehrenfest equations at the critical point.
Moreover, the right hand side of Eq.~(\ref{eq:39a}) is given by
\begin{eqnarray}
\frac{\Delta\zeta}{\Delta\kappa_T}=\left[\frac{\partial P}{\partial T}\Big|_V\right]_{r_+=r_c}
=\frac{3}{2(n+1)r_c}.\label{eq:46a}
\end{eqnarray}
Using Eqs.~(\ref{eq:42a}) and (\ref{eq:46a}), the Prigogine-Defay (PD) ratio $(\Pi)$ \cite{Nieuwenhuizen}
is
\begin{eqnarray}
\Pi=\frac{\Delta C_P\Delta \kappa_T}{Tv(\Delta\zeta)^2}=1.\label{eq:47a}
\end{eqnarray}
Hence, this phase transition at the critical point in the $N$-dimensional
EMPYM black hole is of the second order in both cases of $q=(n+1)/4$ and $q\neq(n+1)/4$.
These results are also consistent with the nature of the liquid-gas phase transition
at the critical point.

\section{closing remarks}
\label{5s}

In this paper we have studied the phase transition and critical behavior of $N$-dimensional AdS black holes
in the Einstein-Maxwell-power-Yang-Mills gravity, where the cosmological constant appears as a dynamical
pressure of the system and its conjugate quantity is the thermodynamic volume of the black hole.
It shows that for the case of $q\neq (n+1)/4$, excepting $q=1/2$, there exists a Van der Waals
phase transition under proper constraint conditions of the charges $C$,$Q$ of Maxwell and Yang-Mills
fields and the dimension $n$. For the case of $q=(n+1)/4$, the P-V criticality and the
small/large black hole phase transition always exist with no constraint condition.

In the case of $q\neq (n+1)/4$, the power Yang-Mills term contributes a positive term or a negative term to the metric function \eq{fr}, which depends on the range of the power exponent $q$. However, the contribution of the YM term in the equation of state \eq{equ:p0} is always positive. Comparing \eq{equ:p0} with the equation of state in Ref.\cite{Kubiznak:2012wp,Gunasekaran:2012dq}, we can find that the first and last terms in \eq{equ:p0} are standard and the middle terms are both positive monotonically decreasing functions of radius $r$, which means the equation of state can at most produce a single minimum and a single maximum. This discussion is consistent with the results we calculated, which shows that there is at most one critical point in the system. This is why no re-entrant behaviour is observed and only the Van der Waals behaviour may exist. For the same reason, we can only observed the Van der Waals behaviour in the case of $q=(n+1)/4$.

We have also calculated the critical exponents at the critical point and found in all cases the
critical exponents coincide with those of the Van der Waals fluid. Finally, both of
the Ehrenfest's equations have been verified to hold at the critical point,
which shows that in resemblance with the liquid-gas phase transition,
the phase transition of the EMPYM black hole at the critical point is of the second order.

{\bf Acknowledgments}

This work was supported by the National Natural Science Foundation of China (NSFC)
under Grant Nos. 11275099 and 11347605. and the most important among all the top
priority disciplines of Zhe Jiang Province under Grant Nos. zx2012000070.

\end{document}